\documentclass[aps,pre,twocolumn,10pt,superscriptaddress,nofootinbib,balancelastpage]{revtex4-1} 
\usepackage[latin1]{inputenc}
\usepackage{amsmath,amssymb}
\usepackage{mathrsfs} 
\usepackage{siunitx}
\usepackage{braket}
\usepackage{calc}
\usepackage{tabularx}
\usepackage{graphicx}
\usepackage{dsfont}
\usepackage{color}
\usepackage{ifthen}
\usepackage{fancyhdr}
\usepackage{appendix}
\usepackage{seqsplit}

\allowdisplaybreaks

\usepackage{etoolbox}
\usepackage{hyperref}
\apptocmd{\UrlBreaks}{\do\/\do\-}{}{}

\usepackage[capitalise]{cleveref}

\newcommand{\ii}{\mathrm{i}}%
\newcommand{\dif}{\mathrm{d}}%

\newcommand{\tdif}[2]{\frac{\dif#1}{\dif#2}}%
\newcommand{\Nabla}{\vec{\nabla}}%
\newcommand{\Rea}{\operatorname{Re}}%
\newcommand{\ZT}[1]{\textquotedblleft#1\textquotedblright}%
 
\newcolumntype{Y}{>{\centering\arraybackslash}X}%
\newcolumntype{Z}{>{\raggedright\arraybackslash}X}%

\newlength{\myl}%
\newcommand{\SUM}[2]{{\setlength{\myl}{\widthof{$\displaystyle\sum_{#1}^{#2}$}*\real{0.5}-\widthof{$\displaystyle\sum$}*\real{0.5}}\sum_{#1}^{#2}\;\hspace{-\the\myl}}}
\newcommand{\INT}[3]{\settowidth{\myl}{$\displaystyle\int_{#1}^{#2}$}{\int_{#1}^{#2}\;\;\;\hspace{-\the\myl}\dif #3}\,}
\newcommand{\TINT}[3]{\settowidth{\myl}{$\int_{#1}^{#2}$}{\int_{#1}^{#2}\!\ifthenelse{\equal{#1#2}{}}{}{\;\;\;\;\hspace{-\the\myl}}\dif #3}\,}%
\newcommand{\EINT}[3]{\settowidth{\myl}{$\int_{#1}^{#2}$}{\int_{#1}^{#2}\;\;\;\,\hspace{-\the\myl}\dif #3}\,}

\newcommand{\rt}{(\vec{r},t)}

\begin{document}
	
\title{Effects of social distancing and isolation on epidemic spreading: a dynamical density functional theory model}

\author{Michael te Vrugt}
\affiliation{Institut f\"ur Theoretische Physik, Center for Soft Nanoscience, Westf\"alische Wilhelms-Universit\"at M\"unster, D-48149 M\"unster, Germany}

\author{Jens Bickmann}
\affiliation{Institut f\"ur Theoretische Physik, Center for Soft Nanoscience, Westf\"alische Wilhelms-Universit\"at M\"unster, D-48149 M\"unster, Germany}

\author{Raphael Wittkowski}
\email[Corresponding author: ]{raphael.wittkowski@uni-muenster.de}
\affiliation{Institut f\"ur Theoretische Physik, Center for Soft Nanoscience, Westf\"alische Wilhelms-Universit\"at M\"unster, D-48149 M\"unster, Germany}

\begin{abstract}		
For preventing the spread of epidemics such as the coronavirus disease COVID-19, social distancing and the isolation of infected persons are crucial. However, existing reaction-diffusion equations for epidemic spreading are incapable of describing these effects. We present an extended model for disease spread based on combining an SIR model with a dynamical density functional theory where social distancing and isolation of infected persons are explicitly taken into account. The model shows interesting nonequilibrium phase separation associated with a reduction of the number of infections, and allows for new insights into the control of pandemics.
\end{abstract}
\maketitle

\section{Introduction}
Controlling the spread of infectious diseases, such as the plague \cite{PolandD1998} or the Spanish flu \cite{Wilton1993}, has been an important topic throughout human history \cite{CliffS2013}. Currently, it is of particular interest due to the worldwide outbreak of the coronavirus disease 2019 (COVID-19) induced by the novel coronavirus SARS-CoV-2 \cite{WuEtAl2020,ZhouEtAl2020,CuiLS2019,WangHHG2020}. The spread of this disease is difficult to control, since the majority of infections are not detected \cite{LiPCSZYS2020}. Due to the lack of vaccines, attempts to control the pandemic have mainly focused on social distancing \cite{Stein2020,FergusonEtAl2020} and quarantine \cite{MizumotoC2020,LauKKMSBK2020}, i.e., the general reduction of social interactions, and in particular the isolation of persons with actual or suspected infection. While political decisions on such measures require a way for predicting their effects, existing theories do not explicitly take them into account. 

In this article, we present a dynamical density functional theory (DDFT) \cite{MarconiT1999,ArcherE2004} for epidemic spreading that allows to model the effect of social distancing and isolation on infection numbers. Our model is based on combining the general idea of a reaction-diffusion DDFT from soft matter physics with the SIR model from theoretical biology. The phase diagram predicted by our model shows that, at parameter values corresponding to certain strengths and ratios of social distancing and self-isolation, the system undergoes a phase transition to a state where the spread of the pandemic is suppressed. For all regions of the phase diagram, the predicted curves have the shape that is observed in real pandemics. Numerically, the inhibition of epidemic spreading is found to be associated with nonequilibrium phase separation, where infected persons accumulate at certain spots (\ZT{leper colonies}). Our results are of high interest for the control of pandemics, since the effects of social distancing and the properties of the disease can be studied separately. Moreover, the observed phase separation effects can also be expected to occur in crowded (bio-)chemical systems, which are governed by similar equations. 

\section{Results}
\subsection{SIR-DDFT model}
A quantitative understanding of disease spreading can be gained from mathematical models \cite{CaiGG2015,LeventhalHNB2015,DeDomenicoGPA2016,GomezSA2018}. A well-known theory for epidemic dynamics is the SIR model \cite{KermackM1927}
\begin{align}
\dot{S} &= - cSI\label{s},\\  
\dot{I} &= cSI - wI\label{i},\\ 
\dot{R} &= wI\label{r},
\end{align}
which has already been applied to the current coronavirus outbreak \cite{Nesteruk2020,SimhaPN2020}. It is a reaction-model that describes the number of susceptible $S$, infected $I$, and recovered $R$ individuals as a function of time $t$. Susceptible individuals get the disease when meeting infected individuals at a rate $c$. Infected persons recover from the disease at a rate $w$. When persons have recovered, they are immune to the disease. 

A drawback of this model is that it describes a spatially homogeneous dynamics, i.e., it does not take into account the fact that healthy and infected persons are not distributed homogeneously in space, even though this fact can have significant influence on the pandemic \cite{ZhongGC2020,WangW2018}. To allow for spatial dynamics, disease-spreading theories such as the SIR model have been extended to reaction-diffusion equations \cite{,ColizzaPV2007,PostnikovS2007,NaetherPS2008,BelikGB2011,WangCWWL2012,BacaerS2005,PengL2009}, where a term $D_\phi\Nabla^2 \phi$ with diffusion constant $D_\phi$ is added on the right-hand side of the dynamical equation for $\phi = S,I,R$.

Reaction-diffusion equations, however, still have the problem that they -- being based on the standard diffusion equation -- do not take into account particle interactions other than the reactions. This issue arises, e.g., in chemical reactions in crowded environments such as inside a cell. In this case, the reactands, which are not pointlike, cannot move freely, which prevents them from meeting and thus from reacting. To get an improved model, one can make use of the fact that the diffusion equation is a special case of DDFT. In this theory, the time evolution of a density field $\rho\rt$ with spatial variable $\vec{r}$ is given by
\begin{equation}
\partial_t\rho = \Gamma \Nabla \cdot\bigg(\rho \Nabla \frac{\delta F}{\delta \rho}\bigg)
\label{ddft}%
\end{equation}
with a mobility $\Gamma$ and a free energy $F$. Note that we have written \cref{ddft} without noise terms, which implies that $\rho\rt$ denotes an ensemble average \cite{ArcherR2004}. The free energy is given by 
\begin{equation}
F = F_{\mathrm{id}} + F_{\mathrm{exc}} + F_{\mathrm{ext}}.
\end{equation}
Its first contribution is the ideal gas free energy
\begin{equation}
F_{\mathrm{id}} = \beta^{-1}\INT{}{}{^d r}\rho(\vec{r},t)(\ln(\rho(\vec{r},t)\Lambda^d) -1),
\end{equation}
corresponding to a system of noninteracting particles with the inverse temperature $\beta$, number of spatial dimensions $d$, and thermal de Broglie wavelength $\Lambda$. If this is the only contribution, \cref{ddft} reduces to the standard diffusion equation with $D = \Gamma \beta^{-1}$. The second contribution is the excess free energy $F_{\mathrm{exc}}$, which takes the effect of particle interactions into account. It is typically not known exactly and has to be approximated. The third contribution $F_{\mathrm{ext}}$ incorporates the effect of an external potential $U_{\mathrm{ext}}\rt$. DDFT can be extended to mixtures \cite{Archer2005,WittkowskiLB2012}, which makes it applicable to chemical reactions. While DDFT is not an exact theory (it is based on the assumption that the density is the only slow variable in the system \cite{EspanolL2009,teVrugtW2019d}), it is nevertheless a significant improvement compared to the standard diffusion equation as it allows to incorporate the effects of particle interactions and generally shows excellent agreement with microscopic simulations. In particular, it allows to incorporate the effects of particle interactions such as crowding in reaction-diffusion equations. This is done by replacing the diffusion term $D\Nabla^2\phi\rt$ in the standard reaction-diffusion model with the right-hand side of the DDFT equation \eqref{ddft} \cite{Lutsko2016,LutskoN2016,LiuL2020}. Thus, given that its equilibrium counterpart, static density functional theory, has already been used to model crowds \cite{MendezKYSCA2018}, DDFT is a very promising approach for the development of extended models for epidemic spreading. However, despite the successes of DDFT in other biological contexts such as cancer growth \cite{AlSaediHAW2018}, protein adsorption \cite{AngiolettiBD2018}, ecology \cite{MartinezCMOL2013}, or active matter \cite{WensinkL2008,WittkowskiL2011,MenzelSHL2016,HoellLM2019,PototskyS2012,WittmannB2016,WittmannMMB2017}, no attempts have been made to apply DDFT to epidemic spreading (or other types of socio-economic dynamics).

We use the idea of a reaction-diffusion DDFT to extend the SIR model given by Eqs.\ \eqref{s}-\eqref{r} to a (probably spatially inhomogeneous) system of \textit{interacting} persons, which compared to existing methods allows the incorporation of social interactions and social distancing. Persons are modelled as diffusing particles that can be susceptible to, infected with, or recovered from a certain disease. Social distancing and self-isolation are incorporated as repulsive interactions. The dynamics of the interacting particles can then be described by DDFT, while reaction terms account for disease transmission and recovery. DDFT describes the diffusive relaxation of an interacting system and is thus appropriate if we make the plausible approximation that the underlying diffusion behavior of persons is Markovian \cite{teVrugt2020} and ergodic \cite{SchindlerWB2019}. Using the Mori-Zwanzig formalism \cite{Mori1965,Zwanzig1960,teVrugtW2019}, one can connect the DDFT model and its coefficients to the dynamics of the individual persons \cite{EspanolL2009,teVrugtW2019d}. The extended model reads
\begin{align}
\partial_tS &= \Gamma_S\Nabla\cdot\bigg( S\Nabla\frac{\delta F}{\delta S}\bigg) - cSI\label{sr},\\  
\partial_tI &= \Gamma_I\Nabla\cdot\bigg( I \Nabla\frac{\delta F}{\delta I}\bigg) +cSI - wI -mI\label{ir},\\ 
\partial_tR &=\Gamma_R\Nabla\cdot\bigg( R \Nabla\frac{\delta F}{\delta R}\bigg) + wI\label{rr}.
\end{align}
Note that we use different mobilities $\Gamma_S$, $\Gamma_I$, and $\Gamma_R$ for the different fields $S$, $I$, and $R$, which allows to model the fact that infected persons, who might be in quarantine, move less than healthy persons. For generality, we have added a term $-mI$ on the right-hand side of \cref{ir} to allow for death of infected persons, which occurs at a rate $m$ (cf.\ SIRD model \cite{ZhuWLGW2019,BergeLMMK2017}). Since we are mainly interested in how fast the infection spreads, we will set $m=0$ in the following. In this case, since the total number of persons is constant, one can easily show that
\begin{equation}
\vec{J} = - \Gamma_S S\Nabla\frac{\delta F}{\delta S} - \Gamma_I I \Nabla\frac{\delta F}{\delta I} - \Gamma_R R \Nabla\frac{\delta F}{\delta R}
\end{equation}
is a conserved current. The ideal gas term $F_{\mathrm{id}}$ in the free energy corresponds to a system of noninteracting persons and ensures that standard reaction-diffusion models for disease spreading \cite{NaetherPS2008} arise as a limiting case. The temperature measures the intensity of motion of the persons. A normal social life corresponds to an average temperature, while the restrictions associated with a pandemic will lead to a lower temperature. Moreover, the temperature can be position-dependent if the epidemic is dealt with differently in different places. The excess free energy $F_{\mathrm{exc}}$ describes interactions. This is crucial here as it allows to model effects of social distancing and self-isolation via a repulsive potential between the different persons. Social distancing is a repulsion between healthy persons, while self-isolation corresponds to a stronger repulsive potential between infected persons and other persons. Thus, we set
\begin{equation}
F_{\mathrm{exc}} = F_{\mathrm{sd}} + F_{\mathrm{si}},\label{ex}
\end{equation}
with $F_{\mathrm{sd}}$ describing social distancing and $F_{\mathrm{si}}$ self-isolation. Note that effects of such a repulsive interaction are not necessarily covered by a general reduction of the diffusivity in existing reaction-diffusion models. For example, if people practice social distancing, they will keep a certain distance (6 feet is recommended \cite{Zhu2020}) in places such as supermarkets, where persons accumulate even during a pandemic, or if people live in crowded environments, as was the case on the ship \ZT{Diamond Princess} \cite{MizumotoC2020}. In our model, in the cases of two particles approaching each other, which even at lower temperatures still happens, repulsive interactions will reduce the probability of a collision and thus of an infection. Existing models can only incorporate this in an effective way as a reduction of the transmission rate $c$, which implies, however, that properties of the disease (How infectious is it?) and measures implemented against it (Do people stay away from each other?) cannot be modelled independently. Furthermore, interactions allow for the emergence of spatio-temporal patterns. The final contribution is the external potential $U_{\mathrm{ext}}$. In general, it allows to incorporate effects of confinement into DDFT. Here, it corresponds to things such as externally imposed restrictions of movement. Travel bans or the isolation of a region with high rates of infection enter the model as potential wells.

The advantage of our model compared to the standard SIR theory is that it allows -- in a way that is computationally much less expensive than \ZT{microscopic} simulations, since the computational cost is independent of the number of persons \cite{MalijevskyA2013} -- to study the way in which different actions affect how the disease spreads. For example, people staying at home corresponds to reducing the temperature, quarantine measures correspond to a strongly repulsive potential between infected an healthy persons, and mass events correspond to attractive potentials.

Specifically, we assume that both types of interactions can be modelled via Gaussian pair potentials, depending on the parameters $C_{\mathrm{sd}}$ and $C_{\mathrm{si}}$ determining the strength and $\sigma_{\mathrm{sd}}$ and $\sigma_{\mathrm{si}}$ determining the range of the interactions. Combining this assumption with a Ramakrishnan-Yussouff approximation \cite{RamakrishnanY1979} for the excess free energy and a Debye-H\"uckel approximation \cite{HansenMD2009} for the two-body correlation, we get the specific \textit{SIR-DDFT model} 
\begin{align}
\begin{split}
\partial_tS &= D_S\Nabla^2 S - \Gamma_S\Nabla\cdot\big(S \Nabla (C_{\mathrm{sd}}K_{\mathrm{sd}}\star (S+R) \\
&\quad\!\:+ C_{\mathrm{si}}K_{\mathrm{si}}\star I)\big) - cS I\label{srspec},
\end{split}\raisetag{1.5em}\\  
\begin{split}
\partial_tI &= D_I\Nabla^2 I - \Gamma_I\Nabla\cdot\big( I\Nabla(C_{\mathrm{si}}K_{\mathrm{si}}\star(S+I+R))\big)\\
&\quad\!\:+cS I- wI\label{irspec},
\end{split}\raisetag{1.5em}\\
\begin{split}
\partial_tR &= D_R\Nabla^2 R - \Gamma_R\Nabla\cdot\big( R\Nabla (C_{\mathrm{sd}}K_{\mathrm{sd}}\star (S+R) \\
&\quad\!\:+ C_{\mathrm{si}}K_{\mathrm{si}}\star I)\big)+ wI\label{rrspec}
\end{split}\raisetag{1.5em}
\end{align}
with the diffusion coefficients $D_\phi=\Gamma_\phi \beta^{-1}$ for $\phi=S,I,R$, the kernels
\begin{align}
K_\mathrm{sd}(\vec{r})&=\exp(-\sigma_\mathrm{sd}\vec{r}^2),\\
K_\mathrm{si}(\vec{r})&=\exp(-\sigma_\mathrm{si}\vec{r}^2),
\end{align}
and the spatial convolution $\star$. A possible generalization is discussed in the Supplementary Information.

\subsection{Disease outbreak}
We perform a linear stability analysis of this model, using a general pair potential, in order to determine whether a homogeneous state with $I=0$, which is always a fixed point, is stable. This provides an analytical criterion for whether a disease outbreak will occur. The full calculation is given in \cref{linear}. In the simple SIR model, the $S$-$R$ plane in phase space (these are the states where everyone is healthy) becomes unstable when $c S_0 > w$, where $S_0$ is the initial number of susceptible persons. Thus, the pandemic cannot break out if persons recover faster than they are able to infect others. A linear stability analysis of the full model, performed under the assumption that the initial number of immune persons $R_0$ is small (which corresponds to a new disease) gives the eigenvalue $\lambda_1 = cS_0 -w - D_I k^2$ with the wavenumber $k$, such that this instability criterion still holds when interactions are present. This means that social distancing cannot stabilize a state without infected persons, and can thus not prevent the outbreak of a disease. As reported in the literature \cite{NaetherPS2008}, the marginal stability hypothesis \cite{DeeL1983,BenJacobDK1985,ArcherRTK2012,ArcherWTK2014,ArcherWTK2016} gives, based on this dispersion, a front propagation speed of $v=2\sqrt{D_{I}(cS_0 - w)}$. (This is shown explicitly in \cref{front}.) However, there are two additional eigenvalues $\lambda_{2/3} = (-D_{j} + j_0 \Gamma_j U_{\mathrm{sd}}\hat{h}_d(k))k^2$ with $j=S,R$ and the Fourier transformed social distancing potential $U_{\mathrm{sd}} \hat{h}_d(k)$ associated with instabilities due to interactions. Front speeds for dispersions of this form have been calculated by \citet{ArcherRTK2012}. If both epidemic and interaction modes are unstable, the fronts might interfere, leading to interesting results depending on their different speeds.

\subsection{Inhibition of epidemic by quarantine}
For a further analysis, we solved Eqs.\ \eqref{srspec}-\eqref{rrspec} numerically. Details on the simulations can be found in \cref{numeric}. The relevant control parameters are $C_{\mathrm{sd}}$ and $C_{\mathrm{si}}$, which control the effects of social interactions that are the new aspect of our model. We assume these parameters to be $\leq 0$, which corresponds to repulsive interactions. We assume $x$ and $t$ to be dimensionless, such that all model parameters can be dimensionless too. 

\begin{figure}[htb]
\centering
\includegraphics[width=\linewidth]{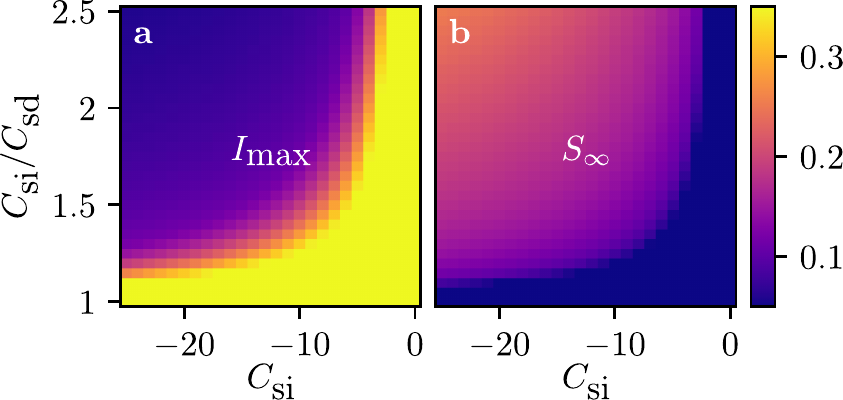}
\caption{\label{fig:phasediagram}\textbf{Phase diagram for the SIR-DDFT model.} The plots show the dependence of (a) the maximal number of infected persons $I_{\mathrm{max}}$ and (b) the final number of susceptible persons $S_\infty$ on the strength of self-isolation $C_{\mathrm{si}}$ and social distancing $C_{\mathrm{sd}}$. A phase boundary is clearly visible.}
\end{figure}

Measures implemented against a pandemic will typically have two aims: reduction of the total number of infected persons, i.e., making sure that the final number of noninfected persons $S_\infty=\lim_{t \to \infty} S(t)$ is large, and reduction of the maximum number of infected persons $I_{\mathrm{max}}$ for keeping the spread within the capacities of the healthcare system. Using parameter scans, we can test whether social distancing and self-isolation can achieve those effects. 

As can be seen from the phase diagrams for the SIR-DDFT model shown in Fig.\ \ref{fig:phasediagram}, there is a clear phase boundary between the upper left corner, where low values of $I_{\mathrm{max}}$ and high values of $S_\infty$ show that the spread of the disease has been significantly reduced, and the rest of the phase diagram, where the disease spreads in essentially the same way as in the model without social distancing. Since all simulations were performed with parameters of $c$ and $w$ that correspond to a disease outbreak in the usual SIR model, this shows that a reduction of social interactions can significantly inhibit epidemic spreading, and that the SIR-DDFT model is capable of demonstrating these effects. The phase boundary shows that, for a reduction of spreading by social measures, two conditions have to be satisfied. First, $|C_{\mathrm{si}}|$ has to be sufficiently large. Second, $|C_{\mathrm{si}}|$ has to be, by a certain amount, larger than $|C_{\mathrm{sd}}|$. Within our physical model of repulsively interacting particles, this arises from the fact that if healthy \ZT{particles} are repelled more strongly by other healthy particles than by infected ones, they will spend more time near infected particles and thus are more likely to be infected themselves. Physically, $|C_{\mathrm{si}}|>|C_{\mathrm{sd}}|$ is thus a very reasonable condition given that infected persons, at least once they develop symptoms, will be isolated more strongly than healthy persons.
\begin{figure}[tb]
\centering
\includegraphics[width=\linewidth]{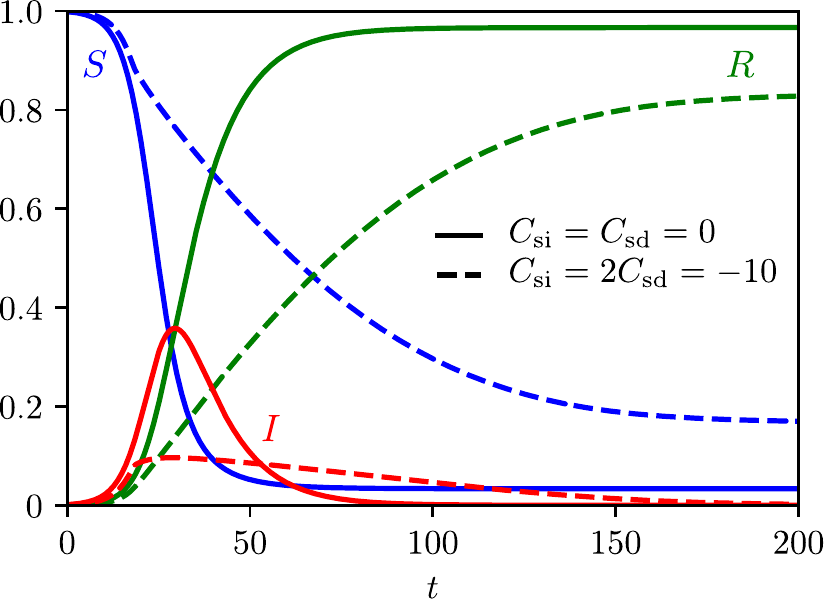}
\caption{\label{fig:average}\textbf{Time evolution of the total number of susceptible, infected, and recovered persons.} The time evolution is shown for no interactions ($C_{\mathrm{si}} = C_{\mathrm{sd}} = 0$) and for interactions with $C_{\mathrm{si}} = 2C_{\mathrm{sd}} = -10$. It can be seen that a reduction of social contacts flattens the curve $I(t)$.}
\end{figure}
Figure \ref{fig:average} shows the time evolution of the total numbers $S(t)$, $I(t)$, and $R(t)$ of susceptible, infected, and recovered persons, respectively, for the cases without interactions (usual SIR model with diffusion) and with interactions (our model). If no interactions are present (i.e., $C_{\mathrm{si}} = C_{\mathrm{sd}} = 0$), $I(t)$ reaches a maximum value of about 0.4 and the pandemic is over at time $t \approx 100$. In the case with interactions (we choose $C_{\mathrm{si}} = 2C_{\mathrm{sd}} = -10$, i.e., parameter values inside the social isolation phase), the maximum is significantly reduced to a value of about 0.1. The final value of $R(t)$, which measures the total number of persons that have been infected during the pandemic, decreases from about 1.0 to about 0.8. Moreover, it takes significantly longer (until time $t \approx 200$) for the pandemic to end. This demonstrates that social distancing and self-isolation have the effects they are supposed to have, i.e., to flatten the curve $I(t)$ in such a way that the healthcare system is able to take care of all cases. Notably, all curves were obtained with the same values of $c$ and $w$, i.e., the properties of the disease are identical. Hence, the observed effect is solely a consequence of social interactions. In the usual SIR model, in contrast, these would be accounted for by modifying $c$, such that they could not be studied separately. The theoretical predictions for the effects of quarantine on the course of $I(t)$ (sharp rise, followed by a bend and a flat curve) are in good qualitative agreement with recent data from China \cite{Hopkins2020,EllyatTL2020}, where strict regulations were implemented to control the COVID-19 spread \cite{LauKKMSBK2020}.

\subsection{Phase separation}
\begin{figure}
\centering
\includegraphics[width=\linewidth]{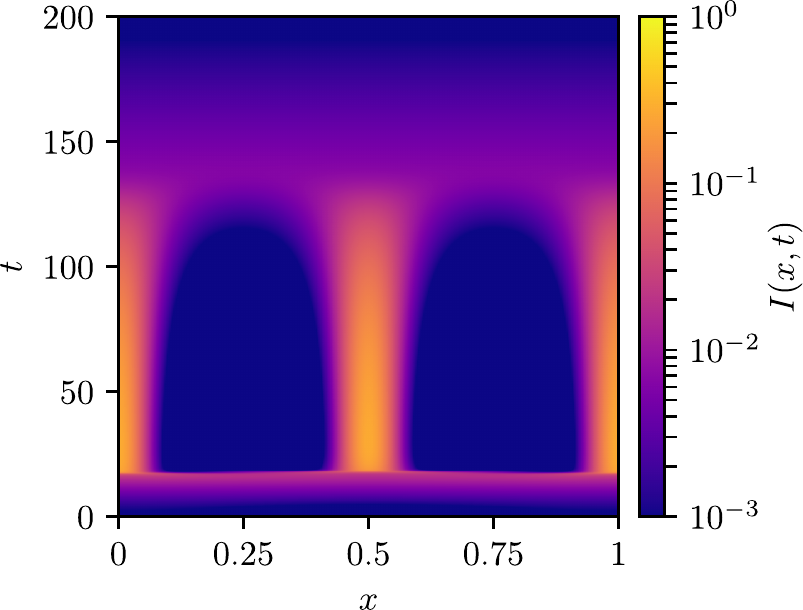}
\caption{\label{fig:spacetime}\textbf{Number of infected persons $\boldsymbol{I(x,t)}$ as a function of space $\boldsymbol{x}$ and time $\boldsymbol{t}$.} During the epidemic spreading, the infected persons self-organize into \ZT{leper colonies}.}
\end{figure}

To explain the observed phenomena, it is helpful to analyze the spatial distribution of susceptible and infected persons during the pandemic. Figure \ref{fig:spacetime} visualizes $I(x,t)$ with $x=(\vec{r})_1$. Interestingly, during the time interval where the pandemic is present, a phase separation can be observed in which the infected persons accumulate at certain spots separated from the susceptible persons. (As this effect is reminiscent of measures that used to be implemented against the spread of leprosy, we refer to these spots as \ZT{leper colonies}.) This phase separation is a consequence of the interactions. Since the formation of leper colonies reduces the spatial overlap of the functions $I(x,t)$ and $S(x,t)$, i.e., the amount of contacts between infected and susceptible persons, the total number of infections decreases significantly and it takes longer until enough persons are immune to stop the pandemic.

The leper colony transition is an interesting type of nonequilibrium phase behavior in its own right. Recall that we have motivated the SIR-DDFT model based on theories for nonideal chemical reactions. It is thus very likely that effects similar to the ones observed here can be found in chemistry. In this case, they would imply that particle interactions can significantly affect the amount of a certain substance that is produced within a chemical reaction, and that such reactions are accompanied by new types of (transient) pattern formation.
\section{Discussion}
In summary, we have presented a DDFT-based extension of the usual models for epidemic spreading that allows to incorporate social interactions, in particular in the form of self-isolation and social distancing. This has allowed us to analyze the effect of these measures on the spatio-temporal evolution of pandemics. Given the importance of the reduction of social interactions for the control of pandemics, the model provides a highly useful new tool for predicting epidemics and deciding how to react to them. Moreover, it shows an interesting phase behavior relevant for future work on DDFT and nonideal chemical reactions. A possible extension of our model is the incorporation of fractional derivatives \cite{MuhammadA2019,QureshiY2019}. Furthermore, enhanced simulations in two spatial dimensions could show interesting pattern formation effects associated with leper colony formation.

\section{Methods}
\subsection{\label{linear}Linear stability analysis}
Here, we perform a linear stability analysis of the extended model given by \cref{sr,ir,rr}. For the excess free energy, we use the combined Ramakrishnan-Yussouff-Debye-H\"uckel approximation as in \cref{srspec,irspec,rrspec}, but now with general two-body potentials $U_{\mathrm{sd}}h_d(x-x')$ for social distancing and $U_{\mathrm{si}}h_i(x-x')$ for self-isolation. In one spatial dimension, we obtain
\begin{align}
\begin{split}
\partial_t S(x,t)& = D_S \partial^2_xS(x,t) - cS(x,t)I(x,t)\\&\quad-\Gamma_S U_{\mathrm{sd}}\partial_x\bigg( S(x,t)\partial_x \INT{}{}{x'}h_d(x-x')\\&\qquad\qquad\qquad\quad(S(x',t)+R(x',t))\bigg)\\
&\quad-\Gamma_S U_{\mathrm{si}}\partial_x\bigg( S(x,t)\partial_x \INT{}{}{x'}h_i(x-x')I(x',t)\bigg),
\end{split}\raisetag{5em}\\
\begin{split}
\partial_t I(x,t)& = D_I \partial^2_xI(x,t) + cS(x,t)I(x,t) - wI(x,t)\\&\quad-\Gamma_{I} U_{\mathrm{si}}\partial_x \bigg(I(x,t)\partial_x \INT{}{}{x'}h_i(x-x')\\&\qquad\qquad\qquad\quad(S(x',t)+I(x',t) +R(x',t))\bigg),    
\end{split}\raisetag{4.5em}\\   
\begin{split}
\partial_t R(x,t) &= D_R \partial^2_xR(x,t) +wI(x,t)\\
&\quad-\Gamma_R U_{\mathrm{sd}}\partial_x\bigg( R(x,t)\partial_x \INT{}{}{x'}h_d(x-x')\\&\qquad\qquad\qquad\quad(S(x',t)+R(x',t))\bigg)\\
&\quad-\Gamma_{R} U_{\mathrm{si}}\partial_x\bigg( R(x,t)\partial_x \INT{}{}{x'}h_i(x-x') I(x',t)\bigg). 
\end{split}\raisetag{5em}
\end{align}
Any homogeneous state with $S=S_0$, $R = R_0$, and $I=0$, where $S_0$ and $R_0$ are constants, will be a fixed point. We consider fields $S = S_0 + \tilde{S}$ and $R = R_0 + \tilde{R}$ with small perturbations $\tilde{S}$ and $\tilde{R}$ and linearize in the perturbations. This results in
\begin{align}
\begin{split}
\partial_t \tilde{S}(x,t) &= D_S \partial^2_x \tilde{S}(x,t) - cS_0I(x,t)\\
&\quad-S_0\Gamma_S U_{\mathrm{sd}}\partial^2_x \INT{}{}{x'}h_d(x-x')(\tilde{S}(x',t)+\tilde{R}(x',t))\\
&\quad-S_0 \Gamma_S U_{\mathrm{si}}\partial^2_x \INT{}{}{x'}h_i(x-x')I(x',t),
\end{split}\raisetag{2em}\\
\begin{split}
\partial_t I(x,t)&= D_I \partial^2_x I(x,t) + cS_0I(x,t) - wI(x,t),    
\end{split}\\
\begin{split}
\partial_t \tilde{R}(x,t) &= D_R \partial^2_x \tilde{R}(x,t) +wI(x,t)\\
&\quad-R_0\Gamma_R U_{\mathrm{sd}}\partial^2_x \INT{}{}{x'}h_d(x-x')(\tilde{S}(x',t)+\tilde{R}(x',t))\\
&\quad-R_0 \Gamma_{R} U_{\mathrm{si}}\partial^2_x \INT{}{}{x'}h_i(x-x')I(x',t).
\end{split}\raisetag{2em}
\end{align}

We now drop the tilde and make the ansatz $S = S_1 \exp(\lambda t - \ii k x)$, $I = I_1 \exp(\lambda t - \ii k x)$, and $R = R_1 \exp(\lambda t - \ii k x)$. This gives the eigenvalue equation
\begin{widetext}
\begin{align}
\lambda
\begin{pmatrix}
S_1\\
I_1\\
R_1\\
\end{pmatrix}
= 
\begin{pmatrix}
- D_S k^2 + S_0\Gamma_S U_{\mathrm{sd}}\hat{h}_d(k)k^2 & S_0 \Gamma_S U_{\mathrm{si}}\hat{h}_i(k)k^2 - cS_0 & S_0\Gamma_S U_{\mathrm{sd}}\hat{h}_d(k)k^2\\
0 & - D_I k^2 + cS_0  - w & 0 \\
R_0\Gamma_R U_{\mathrm{sd}}\hat{h}_d(k)k^2 & w + R_0\Gamma_R U_{\mathrm{si}}\hat{h}_i(k)k^2 & - D_R k^2 + R_0\Gamma_R U_{\mathrm{sd}}\hat{h}_d(k)k^2
\end{pmatrix}\!\!
\begin{pmatrix}
S_1\\
I_1\\
R_1\\
\end{pmatrix}.
\label{matrix}%
\end{align}
\end{widetext}
Here, $\hat{h}_d(k)$ and $\hat{h}_i(k)$ are the Fourier transforms of $h_d(x-x')$ and $h_i(x-x')$, respectively. The corresponding characteristic polynomial reads
\begin{equation}
\begin{split}
(-\lambda - D_I k^2 + cS_0 - w)&\\
((- D_S k^2 + S_0\Gamma_S U_{\mathrm{sd}}\hat{h}_d(k)k^2 - \lambda))&\\
(- D_R k^2 + R_0\Gamma_R U_{\mathrm{sd}}\hat{h}_d(k)k^2 - \lambda))&\\
- S_0R_0 k^4U_{\mathrm{sd}}^2\hat{h}_d^2(k)\Gamma_S\Gamma_R) &= 0.
\label{characteristic}%
\end{split}
\end{equation}
Rather than solving this third-order polynomial in $\lambda$ exactly, we consider the limit of long wavelengths. For $k=0$, which corresponds to the usual SIR model given by Eqs.\ \eqref{s}-\eqref{r} in the main text if we assume $k^2 \hat{h}_d(k)=0$, \cref{characteristic} simplifies to
\begin{equation}
(-\lambda + cS_0 - w)\lambda^2 = 0,
\end{equation}
which has the solutions
\begin{align}
\lambda_1 &= c S_0 - w,\\
\lambda_2 &= 0.
\end{align}
This means that the epidemic will start growing when $c S_0 > w$, since in this case there is a positive eigenvalue. When interpreting this result, one should take into account that, since a susceptible person that has been infected cannot become susceptible again, the system will, after a small perturbation, not go back to the same state as before even if $w > c S_0$. Actually, we have tested the linear stability of the $S$-$R$ plane in phase space, and the fact that any parameter combination of $S_0$ and $R_0$ with $I=0$ is a solution of the SIR model is reflected by the existence of the eigenvalue $\lambda = 0$ with algebraic multiplicity 2 (a perturbation within the $S$-$R$ plane will obviously not lead to an outbreak).

Next, we consider the case $k \neq 0$, but assume that we can neglect the term $S_0R_0 k^4U_{\mathrm{sd}}^2\hat{h}_d^2(k)\Gamma_S\Gamma_R$ in \cref{characteristic}. This corresponds to assuming either $R_0 = 0$ (i.e., we consider the begin of an outbreak of a new disease that no one is yet immune against) or small $k$ (such that terms of order $k^4$ can be neglected). Then, \cref{characteristic} gives
\begin{equation}
\begin{split}
(-\lambda - D_I k^2 + cS_0 - w)&\\
(- D_S k^2 + S_0\Gamma_S U_{\mathrm{sd}}\hat{h}_d(k)k^2 - \lambda)&\\
(- D_R k^2 + R_0\Gamma_R U_{\mathrm{sd}}\hat{h}_d(k)k^2 - \lambda)&= 0.
\end{split}
\end{equation}
We can immediately read off the solutions
\begin{align}
\lambda_1 &= c S_0 - w - D_I k^2,\\
\lambda_2 &= - D_S k^2 + S_0\Gamma_S U_{\mathrm{sd}}\hat{h}_d(k)k^2,\label{lambda2}\\
\lambda_3 &=- D_R k^2 + R_0\Gamma_R U_{\mathrm{sd}}\hat{h}_d(k)k^2.\label{lambda3}
\end{align}
The result for $\lambda_1$ shows that the initial state still becomes unstable for $c S_0 > w$, i.e., the interactions cannot stabilize a state without infected persons that would be unstable otherwise. The eigenvalues $\lambda_2$ and $\lambda_3$, which were 0 in the long-wavelength limit, now describe the dispersion due to interparticle interactions that may lead to instabilities not related to disease outbreak.
\subsection{\label{front}Front speed}
For determining the propagation speed of fronts, we can use the marginal stability hypothesis \cite{DeeL1983,BenJacobDK1985,ArcherRTK2012,ArcherWTK2014,ArcherWTK2016}. We transform to the co-moving frame that has velocity $v$ and assume that the growth rate in this frame is zero at the leading edge. Thereby, we obtain for a general dispersion $\lambda(k)$ the equations
\begin{align}
\ii v + \tdif{\lambda}{k} &= 0 \label{marginal1},\\
\Rea(\ii vk + \lambda)&= 0\label{marginal2}.
\end{align}
These equations can be solved for the complex wavenumber $k = k_{\mathrm{re}} + \ii k_{\mathrm{im}}$ and the  velocity $v$. For the dispersion $\lambda_1 = c S_0 - w - D_I k^2$ (we are interested in instabilities associated with infections), \cref{marginal1,marginal2} lead to
\begin{align}
\ii v - 2 \ii D_I k_{\mathrm{im}} &=0,\\
- 2 D_I k_{\mathrm{re}} &=0,\\
- v k_{\mathrm{im}} + c S_0 -w - D_I(k_{\mathrm{re}}^2 - k_{\mathrm{im}}^2) &=0.
\end{align}
The solution of these equations is 
\begin{align}
k_{\mathrm{re}} &= 0,\\
k_{\mathrm{im}} &= \pm \sqrt{\frac{c S_0 -w}{D_I}},\\
v &= 2\sqrt{D_I(cS_0 - w)},
\end{align}
which is in agreement with results from the literature \cite{NaetherPS2008}. Front speeds for dispersions of the form \eqref{lambda2} and \eqref{lambda3} can be found in Ref.\ \cite{ArcherRTK2012}.

\subsection{\label{numeric}Numerical analysis}
The calculation was done in one spatial dimension on the domain $x \in [0,1]$ with periodic boundary conditions, using an explicit finite-difference scheme with step size $\dif x = 0.005$ (individual simulations) or $\dif x =0.01$ (parameter scan) and adaptive time steps. As an initial condition, we use a Gaussian peak with amplitude $1$ and variance $50^{-2}$ centered at $x=0.5$ for $S(x,0)$, $I(x,0)=0.001 S(x,0)$, and $R(x,0)=0$. Since the effect of the parameters $c$ and $w$ on the dynamics is known from previous studies of the SIR model, we fix their values to $c=1$ and $w=0.1$ to allow for an outbreak. Moreover, we set $\Gamma_S = \Gamma_I = \Gamma_R = 1$, $D_S = D_I = D_R = 0.01$, and $\sigma_{\mathrm{sd}}=\sigma_{\mathrm{si}}=100$.

\section*{Acknowledgements}
R.W.\ is funded by the Deutsche Forschungsgemeinschaft (DFG, German Research Foundation) -- WI 4170/3-1.

\nocite{apsrev41Control}
\bibliographystyle{apsrev4-1}
\bibliography{control,refs}

\end{document}